\documentclass[10pt,twocolumn]{article}

\usepackage[utf8]{inputenc}
\usepackage{amsmath}
\usepackage{amssymb}
\usepackage{titlesec}
\usepackage{natbib}
\usepackage{booktabs}
\usepackage{graphicx}
\usepackage[colorlinks=false]{hyperref}
\usepackage[format=plain,labelfont=it]{caption}
\usepackage[left=1.5cm,right=1.5cm,top=2cm,bottom=2cm]{geometry}

\usepackage{tikz}
\usetikzlibrary{arrows}
\usetikzlibrary{shapes}
\usetikzlibrary{decorations.pathmorphing}
\usetikzlibrary{decorations.pathreplacing}
\usetikzlibrary{decorations.shapes}
\usetikzlibrary{decorations.text}
\usetikzlibrary{positioning, shapes}

\tikzset{
block/.style = {draw, fill=white, rectangle, minimum height=3em, minimum width=4.5em},
tmp/.style  = {coordinate}, 
sum/.style= {draw, fill=white, circle, node distance=2cm},
input/.style = {coordinate},
output/.style= {coordinate}
pinstyle/.style = {pin edge={to-,thick,black}}
}

\titleformat{\section}{\centering\normalfont\scshape}{\arabic{section}.}{5pt}{}
\titleformat{\subsection}{\normalfont\it}{\arabic{section}.\arabic{subsection}}{5pt}{}
\titleformat{\subsubsection}{\normalfont\it}{\arabic{section}.\arabic{subsection}.\arabic{subsubsection}}{5pt}{}

\newcommand\infoFootnote[1]{%
  \begingroup
  \renewcommand\thefootnote{}\footnote{#1}%
  \addtocounter{footnote}{-1}%
  \endgroup}

\newcommand{\R}{\mathbb{R}} 
\newcommand{\N}{\mathbb{N}}

\newcommand{\Xc}{\mathcal{X}}

\newcommand{\Sc}{\mathcal{S}}
\newcommand{\Nc}{\mathcal{N}}
 
\newcommand{\Tc}{\mathcal{T}}

\newcommand{\Uc}{\mathcal{U}} 
\newcommand{\Fc}{\mathcal{F}}

\newcommand{\Zc}{\mathcal{Z}}

\newcommand{\ub}{\boldsymbol{u}}
\newcommand{\vb}{\boldsymbol{v}}
\newcommand{\xb}{\boldsymbol{x}}
\newcommand{\Ab}{\boldsymbol{A}}
\newcommand{\bb}{\boldsymbol{b}}
\newcommand{\cb}{\boldsymbol{c}}
\newcommand{\Bb}{\boldsymbol{B}}
\newcommand{\Nb}{\boldsymbol{N}}

\newcommand{\Gb}{\boldsymbol{G}}

\newcommand{\Ib}{\boldsymbol{I}}

\newcommand{\Pb}{\boldsymbol{P}}
\newcommand{\Qb}{\boldsymbol{Q}}
\newcommand{\Rb}{\boldsymbol{R}}

\newcommand{\fb}{\boldsymbol{f}}
\newcommand{\gb}{\boldsymbol{g}}

\newcommand{\zerob}{\boldsymbol{0}}

\newcommand{\kappab}{\boldsymbol{\kappa}}

\newtheorem{thm}{Theorem}
\newtheorem{assum}{Assumption}

\newtheorem{defn}{Definition}

\newtheorem{lem}[thm]{Lemma}
\newtheorem{exmp}{Example}

\newtheorem{pf}{Proof}

\title{\vspace{-2mm}\bf Convex NMPC reformulations for a special class of nonlinear multi-input systems with application to rank-one bilinear networks}
\author{Manuel Kl\"adtke and Moritz Schulze Darup\vspace{2mm}}
\date{}

\begin{document}

\maketitle

\textbf{\textit{Abstract}.} {\bf We show that a special class of (nonconvex) NMPC problems admits an exact solution by reformulating them as a finite number of convex subproblems, extending previous results to the multi-input case. Our approach is applicable to a special class of input-affine discrete-time systems, which includes a class of bilinear rank-one systems that is considered useful in modeling certain controlled networks. We illustrate our results with two numerical examples, including the aforementioned rank-one bilinear network.}
\infoFootnote{M. Kl\"aedtke and M. Schulze Darup are with the \href{https://rcs.mb.tu-dortmund.de/}{Control and~Cyberphysical Systems Group}, Faculty of Mechanical Engineering, TU Dortmund University, Germany. E-mails:  \href{mailto:manuel.klaedtke@tu-dortmund.de}{\{manuel.klaedtke, moritz.schulzedarup\}@tu-dortmund.de}. \vspace{0.5mm}}
\infoFootnote{\hspace{-1.5mm}$^\ast$This paper is a \textbf{preprint} of a contribution to the 22nd World Congress of the International Federation of Automatic Control 2023. The DOI of the original paper is \href{https://doi.org/10.1016/j.ifacol.2023.10.1321}{10.1016/j.ifacol.2023.10.1321}.}

\section{Introduction}

Model predictive control (MPC) is a well-studied control approach that is particularly valued for its ability to handle multi-input-multi-output systems and to incorporate system constraints directly into the control design. However, as an optimization-based control approach, MPC inherits not only benefits but also challenges associated with the underlying optimization problems. Especially the great watershed in optimization between convexity and non-convexity \citep{Rockafellar1993} is typically associated with linear and nonlinear MPC (NMPC), respectively, since the latter incorporates the nonlinear system dynamics as equality constraints of a (hence nonconvex) optimization problem. Because convexity is the crucial property that allows one to efficiently obtain not only a locally but also a globally optimal solution of the optimization problem, one usually has to be satisfied with locally optimal (but globally suboptimal) solutions in the context of NMPC. However, although quite rare, there are cases where convex reformulations of normally nonconvex NMPC problems can be derived, usually involving very strict assumptions such as a limited prediction horizon \citep{Lautenschlager2015} or a small class of applicable nonlinear systems \citep{Klaedtke2022}.

In this work, we contribute to the effort of convex NMPC reformulations by focusing on a special class of discrete-time input-affine systems similar to \cite{Klaedtke2022}, but extend these results to multi-input systems. We show that the original nonconvex NMPC problem can be split into a finite number of convex subproblems by appropriately partitioning the constrained state-space, which allows to compute an exact global solution by enumeration of these subproblems. While these results are again accompanied with heavy restrictions on the type of system, shape of constraint sets, and also choice of stage costs, we demonstrate that the applicable class of systems encompasses a subclass of bilinear systems that can be used to model certain types of networks such as traffic or biochemical networks \citep{GHOSH2016}.

The paper is organized as follows. We summarize basics on NMPC and present the special class of nonlinear systems to be discussed together with a useful system transformation in Section~\ref{sec:basics}. In Section~\ref{sec:Convex_NMPC}, we show how to transform the given constraints and choose the step costs and terminal ingredients of the nonconvex optimal control problem (OCP) to yield a finite number of convex subproblems that facilitate the exact solution of the original problem. We demonstrate our results with two numerical examples in Section~\ref{sec:examples}, including an example of a rank-one bilinear network mentioned above.

\section{Basics on NMPC, system specification, and system transformation} \label{sec:basics}

We first recall some basics on NMPC for a general nonlinear discrete-time system
\begin{equation}
\label{eq:generalNonlinearSystem}
\xb(k+1)=\fb(\xb(k),\ub(k))
\end{equation}
with state and input constraints
\begin{equation}
\label{eq:constraints}
\xb(k) \in \Xc \subset \R^n \qquad \text{and} \qquad \ub(k) \in \Uc\subset \R^m
\end{equation}
and then specify the special system class of interest. 

\subsection{Nonlinear model predictive control}
Applying NMPC to a nonlinear system with constraints like \eqref{eq:generalNonlinearSystem} involves solving an OCP of the form 
\begin{align}
\label{eq:NMPC}
V(\xb):= \!\!\!\! \min_{\substack{\hat{\xb}(0),\dots,\hat{\xb}(N),\\\hat{\ub}(0),\dots,\hat{\ub}(N-1)}} \!\!\!\!\!\!\!\!\!\!&\,\,\,\,\,\,\,\,\,\,  \varphi(\hat\xb(N))+\sum_{i=0}^{N-1} \ell(\hat{\xb}(i),\hat{\ub}(i))\!\!\!\!\!\!\!\!\!\!\!\!\!\span \span  \\
\nonumber
\text{s.t.} \qquad   \hat{\xb}(0)&=\xb, \\
\nonumber
\hat{\xb}(i+1)&=\fb(\hat{\xb}(i),\hat{\ub}(i)) &&\forall i \in \{0,\dots,N-1\}, \\
 \nonumber
 (\hat{\xb}(i),\hat{\ub}(i)) & \in \Xc \times \Uc &&\forall i \in \{0,\dots,N-1\},\\
 \hat\xb(N) & \in \Tc \nonumber  
\end{align}
in each time step $k$ for the current state $\xb=\xb(k)$ and applying the first step $\hat{\ub}^\ast(0)$ of the optimal control sequence to the system to obtain closed-loop control. Here, $N\in \N$ is the prediction horizon and $\ell: \R^n \times \R^m \to \R$ denotes the stage cost. The terminal cost $\varphi: \R^n \to \R$ and terminal constraint set $\Tc$ are usually constructed in a way that enforces stability guarantees of the closed-loop system \citep{Mayne2000}. 
Crucially, the nonlinear system dynamics \eqref{eq:generalNonlinearSystem} occur as equality constraints in \eqref{eq:NMPC}, which renders the OCP nonconvex even if the cost functions $\ell, \varphi$ and constraint sets $\Xc, \Uc, \Tc$ are convex.

\subsection{System specification}

In this work, we consider input-affine multi-input systems
\begin{align}
    \xb(k+1) &= \Ab \xb(k)+\sum_{i=1}^{m}\bb_ig_i(\xb(k))\ub_i \label{eq:nonlinearSystem} \\
    &= \Ab \xb(k)+\Bb \Gb(\xb)\ub \label{eq:standard_FL}
\end{align}
where $\Ab \in \R^{n\times n}$, $\bb_i \in \R^n$, $\Bb := \begin{pmatrix}                \bb_1 & \hdots & \bb_m            \end{pmatrix}$, $g_i:\R^n \rightarrow \R$, and $\Gb(\xb):=\mathrm{diag}\left(g_1(\xb(k)), ..., g_m(\xb(k))\right)$. 
The first summand in \eqref{eq:nonlinearSystem} signifies that the unforced dynamics of the system are linear, and the diagonal structure of $\Gb(\xb)$ makes this a very natural extension of the system class considered in \cite{Klaedtke2022} to multi-input systems. We will discuss some difficulties encountered when considering non-diagonal $\Gb(\xb)$ in Section~\ref{sec:Constraint_Trafo}, but a deeper analysis and extension to that case will be left open for future work. All further assumptions that are not captured by the structure of \eqref{eq:nonlinearSystem} are summarized below.
\begin{assum}\label{assum:characterization}
    We assume that
    \begin{equation}
        \Xc = \bigcup_{j=1}^s \Xc_j, \quad \text{with}\: \Xc_j\:\text{convex}\: \forall j \in \{1, ..., s\}, \label{eq:Xpartition}
    \end{equation}
    i.e., $\Xc$ can be decomposed into $s\in \N$ convex (and not necessarily disjoint) subsets $\Xc_j$, which have the following properties. For every $j \in \{1,\dots,s\}$, we either have
\begin{subequations}
\label{eq:concaveConvex}
\begin{align}
  \!\!\!g_i(\xb)&\geq 0, \,\,\, g_i(\eta \xb + (1-\eta)\hat{\xb}) \geq \eta  g_i(\xb) \!+\! (1-\eta)g_i(\hat{\xb})\!\!\!\\
  \nonumber
  \text{or} & \\
  \!\!\!g_i(\xb)&\leq 0, \,\,\, g_i(\eta \xb + (1-\eta)\hat{\xb}) \leq \eta  g_i(\xb) \!+\! (1-\eta)g_i(\hat{\xb})\!\!\!
\end{align}
\end{subequations}
for every $\xb,\hat{\xb} \in \Xc_j$, every $g_i(\xb)$ with $i \in \{1, ..., m\}$, and every $\eta \in (0,1)$. Finally, we assume box constraints containing the origin for each input, i.e., 
$\Uc:=[\underline{u}_1,\overline{u}_1]\times \hdots \times [\underline{u}_m,\overline{u}_m]$ with $\underline{u}_i< 0 < \overline{u}_i$ for all $i \in \{1, ..., m\}$.
\end{assum}
Note that unlike \cite{Klaedtke2022, NMPC2021Exact}, we do not assume $g_i(\zerob)\neq 0$ for any of the functions $g_i(\xb)$, nor do we assume $\zerob \in \mathrm{int}(\Xc_j)$ for any of the convex subsets $\Xc_j$, since this assumption is often not satisfied, in particular for the class of bilinear rank-one networks that we show in Example~\ref{exmp:bilinear} of Section~\ref{sec:examples}. 
Undoubtedly, these assumptions are still very restrictive, but they can be easily interpreted and include the bilinear system classes in \cite{NMPC2021Exact} as well as all system classes analyzed in \cite{Klaedtke2022} as single-input cases. The results of this work can therefore be interpreted as another step towards extending the class of systems to which that approach is applicable.

\subsection{System transformation}

Throughout this paper, we will often alternatively consider the associated linear multi-input system
\begin{equation}
    \xb(k+1)=\Ab \xb(k)+\Bb \vb(k) \label{eq:FL_system}
\end{equation}
with artificial input $\vb \in \R^m$, which is related to \eqref{eq:nonlinearSystem} via
$$
    \vb = \Gb(\xb) \ub.
$$
However, since the inverse $\Gb^{-1}(\xb)$ may not exist for all $\xb \in \Xc$, we define the set
$$
    \Xc^\circ = \{\xb\in\Xc |\det(\Gb(\xb))\neq 0\},
$$
on which we can equivalently state 
$$
    \ub = \Gb^{-1}(\xb)\vb \quad \text{for}\quad  \xb \in \Xc^\circ.
$$
Technically, this approach is based on feedback linearization, where a coordinate transformation and (nonlinear) feedback are used to transform a nonlinear system into an equivalent linear one. Readers familiar with this concept may recognize \eqref{eq:standard_FL} as a system structure where the coordinate transformation is not necessary (or, alternatively, may have already happened, i.e., $\xb$ denotes these new coordinates instead of the original ones). Since feedback linearization is a broad concept, only a tiny part of which we will use in this paper, we will not formally recall it here, but instead refer to \cite{Isidori1995} for the continuous-time case, \cite{Soroush1992} for the discrete-time case, and \cite{Klaedtke2022} for an application to a similar but single-input system class. 

The purpose of introducing \eqref{eq:FL_system} is to replace the troublesome nonlinear equality constraints in \eqref{eq:NMPC} with linear ones. However, in doing so, one must also choose the stage cost $\ell(\xb, \ub)=\hat\ell(\xb, \vb=\Gb(\xb)\ub)$ with respect to the artificial input $\vb$ and also introduce alternative constraints for $\vb$ such that their satisfaction implies $\ub\in\Uc$. How to handle these new ingredients in such a way as to yield convex subproblems is the focus of the next section. Finally, it should be noted that the use of feedback linearization for optimal control has of course been studied in the past (e.g., in \cite{Bacic2002, Gao2012}). The distinctiveness of this work and the preceding \cite{NMPC2021Exact, Klaedtke2022} compared to earlier work is the treatment of state and input constraints leading to the convex subproblems mentioned earlier. This, of course, comes at the price of a very restricted class of systems, but one that has already been and continues to be extended by quite a bit.

\section{Convex NMPC reformulations}\label{sec:Convex_NMPC}

Using the associated linear system \eqref{eq:FL_system} and its artificial input $\vb$, we can state the OCP 
\begin{align}
\label{eq:still_NMPC}
V(\xb):= \!\!\!\! \min_{\substack{\hat{\xb}(0),\dots,\hat{\xb}(N),\\\hat{\vb}(0),\dots,\hat{\vb}(N-1)}} \!\!\!\!\!\!\!\!\!\!&\,\,\,\,\,\,\,\,\,\,  \varphi(\hat{\xb}(N))+\sum_{i=0}^{N-1} \hat\ell(\hat{\xb}(i),\hat{\vb}(i))\!\!\!\!\!\!\!\!\!\!\!\!\!\span  \\
\nonumber
\text{s.t.} \qquad   \hat{\xb}(0)&=\xb, \\
\nonumber
\hat{\xb}(i+1)&=\Ab\hat\xb(i)+\Bb\hat\vb(i) &\forall i \in \{0,\dots,N\!-1\}, \\
 \nonumber
 \begin{pmatrix}
     \hat{\xb}(i) \\
     \hat{\vb}(i)
 \end{pmatrix} & \in \Zc &\forall i \in \{0,\dots,N-1\},\\
 \hat{\xb}(N) &\in \Tc\nonumber,
\end{align}
solve it for $\xb = \xb(k)$ in every time step $k$, and use the first element $\hat\vb^\ast(0)$ of the optimal artificial input sequence to compute the actual optimal input $\ub^\ast(k)=\Gb^{-1}(\xb)\hat\vb^\ast(0)$, which is then applied to the system. However, some questions need to be answered regarding the new stage cost $\hat\ell(\xb,\vb)$, equality constraints \eqref{eq:FL_system}, and (mixed) constraint set $\Zc$. In the following, we will discuss the shape of $\Zc$ that is implied by our assumptions on $\Uc$, and its convex decomposition. Furthermore, we show that $\Zc$ harmonizes very well with our system transformation in the sense that it constrains $\vb$ for $\xb\in\Xc\setminus\Xc^\circ$ in such a way that the dynamics of \eqref{eq:nonlinearSystem} and \eqref{eq:FL_system} implicitly remain consistent, even though the inverse $\Gb^{-1}(\xb)$ does not exist for these $\xb$. Regarding the stage cost, we will discuss how to choose $\hat\ell(\xb,\vb)$ for our convex decomposition as well as its effect on $\ell(\xb,\ub)$, and a simple interpretation of their relation. The choice of terminal ingredients and problems regarding the construction of $\Tc$ will only briefly be discussed, however, we give reference to related earlier work and show examples in Section~\ref{sec:examples}. Finally, we show how to solve \eqref{eq:still_NMPC} via its convex subproblems and give some references regarding the computational complexity.

\subsection{Constraint transformation}\label{sec:Constraint_Trafo}

In order to achieve 
\begin{equation}
    (\xb, \ub)\in \Xc\times\Uc \iff \begin{pmatrix}
     \xb(i) \\
     \vb(i)
 \end{pmatrix} \in \Zc, \label{eq:equivalentConstraints}
\end{equation}
we first note that a simple relation is given via 
$$
    \ub=\Gb^{-1}(\xb)\vb\in\Uc \quad \forall \xb \in \Xc^\circ.
$$
To fill the gaps given by $\xb\in\Xc\setminus\Xc^\circ$, we need to investigate when $\Gb(\xb)$ becomes singular and what this implies for the relation between $\vb$ and $\ub$ as well as between \eqref{eq:nonlinearSystem} and \eqref{eq:FL_system}. Given the diagonal structure of $\Gb(\xb)$, it becomes singular if and only if one or more of the functions $g_i(\xb)$ are equal to zero. In this case, the corresponding inputs $\ub_i$ do not affect the system. Moreover, we then have $\vb_i=g_i(\xb)\ub_i=0$, so the corresponding $\vb_i$ cannot be chosen freely, but must be constrained to zero, since otherwise the system dynamics implied by \eqref{eq:FL_system} are inconsistent with the actual dynamics given by \eqref{eq:nonlinearSystem}. Adding such a conditional equality constraint to the OCP is a nontrivial task and may lead to many edge cases making its solution intractable. However, before handling those constraints, we will first formally define the relation between $\ub$ and $\vb$ for all $\xb\in\Xc$. 
\begin{defn}\label{def:Rel_vu}
    Denote as $\Sc(\xb)$ the index set of functions $g_i(\xb)$ being equal to zero for a given state $\xb$, i.e.,
$$
    \Sc(\xb) := \{i \in \{1, ..., m\}|g_i(\xb)=0\},
$$
and its complement $\Nc(\xb)$ is defined in a similar way. Omitting the arguments of $\Sc(\xb)$ and $\Nc(\xb)$ for notational convenience, $\vb_\Sc$, $\ub_\Sc$, $\vb_\Nc$, $\ub_\Nc$ can be constructed from the rows corresponding to $\Sc$ and $\Nc$, respectively, and $\Gb_\Nc(\xb)=\text{diag}(\gb_\Nc(\xb))$ as the diagonal matrix of nonzero $g_i(\xb)$. The relation between $\vb$ and $\ub$ is given by
\begin{align*}
    \vb_\Sc &= \zerob\\
    \vb_\Nc &= \Gb_\Nc(\xb)\ub_\Nc
\end{align*}
with nonsingular $\Gb_\Nc(\xb)$ for all $\xb\in\Xc$. The inputs $\ub_\Sc$ can be chosen arbitrarily (but should align with $\Uc$).
\end{defn}
 The $\ub_\Sc$ can be chosen arbitrarily, since they do not affect the system, and we will simply choose $\ub_\Sc=\zerob$ to avoid unnecessary control effort, which is also a sensible choice regarding positive definiteness of $\hat\ell(\xb, \vb)$ and $\ell(\xb, \ub)$ as will be shown in Lemma~\ref{lem:posdef_ell}.

Given this relation, we can now simply define $\Zc$ as
\begin{equation}
    \!\Zc\! :=\!\!  \left\{ \left.\!\!\begin{pmatrix}
\xb \\
\vb
\end{pmatrix}\!\! \in\! \R^{n+m} \right| \xb\! \in\! \Xc ,\, \Gb_\Nc^{-1}(\xb)\vb_\Nc \!\in\! \Uc,\, \vb_\Sc = \zerob \right\}, \!\! \label{eq:setZ}
\end{equation}
and given the partitions of $\Xc$ in \eqref{eq:Xpartition}, we can equivalently construct the corresponding partitions
$$
    \Zc_j :=  \left\{ \left.\begin{pmatrix}
\xb \\
\vb
\end{pmatrix} \in \R^{n+m} \right|\, \xb \in \Xc_j ,\, \Gb_\Nc^{-1}(\xb)\vb_\Nc \in \Uc,\, \vb_\Sc = \zerob \right\}, 
$$
that also satisfy
$$
    \Zc = \bigcup_{j=1}^s \Zc_j.
$$
This partitioning is important for multiple reasons, which are related to the Assumption~\ref{assum:characterization} made for $g_i(\xb)$ on every $\Xc_j$ and hence also related to $\Zc_j$. By the assumption that $\Uc$ takes the shape of box constraints and $\Gb(\xb)$ is diagonal, we can equivalently specify $\Gb_\Nc^{-1}(\xb)\ub_\Nc \in \Uc$ as
\begin{equation}
    \underline{\ub}_i\leq\frac{\vb_i}{g_i(\xb)}\leq \overline{\ub}_i \quad \forall i \in \Nc \label{eq:ineq_constrain}
\end{equation}
and rearrange these inequalities, since each $g_i(\xb)$ is either nonnegative or nonpositive on a single $\Xc_j$, i.e., the denominator of \eqref{eq:ineq_constrain} does not change its sign. 
This either yields
\begin{equation}
    g_i(\xb)\underline{\ub}_i\leq\vb_i\leq g_i(\xb)\overline{\ub}_i \quad \text{for}\; i\in\Nc \label{eq:rearranged_constraints_1}
\end{equation}
if $g_i(\xb) > 0$ for all $\xb \in \Xc_j$ or
\begin{equation}
    g_i(\xb)\overline{\ub}_i\leq\vb_i\leq g_i(\xb)\underline{\ub}_i \quad \text{for}\; i\in\Nc\label{eq:rearranged_constraints_2}
\end{equation}
if $g_i(\xb)< 0$ for all $\xb \in \Xc_j$ and exactly one of these cases occurs for every $j\in\{1, ..., s\}$ and every $g_i(\xb)$. Note that, while the fraction in \eqref{eq:ineq_constrain} is only defined for $g_i(\xb)\neq 0$, the rearranged inequalities \eqref{eq:rearranged_constraints_1} and \eqref{eq:rearranged_constraints_2} can also be evaluated for $g_i(\xb)=0$, in which case both inequalities change into the equality $\vb_i=0$. Looking back at Definition~\ref{def:Rel_vu}, this is exactly the condition for $\vb$ needed to keep the systems \eqref{eq:nonlinearSystem} and \eqref{eq:FL_system} consistent. This observation can be used to equivalently reformulate the partitions as
\begin{align}
\label{eq:Z_without_eq}
\Zc_j :=  \biggl\{&\begin{pmatrix}
\xb \\
\vb
\end{pmatrix} \in \R^{n+m} \bigg|\, \xb \in \Xc_j ,\\ &g_i(\xb)\underline{\ub}_i\leq\vb_i\leq g_i(\xb)\overline{\ub}_i\quad \text{if}\: g_i(\xb)\leq 0 \: \text{on}\: \Xc_j \nonumber\\
&\text{or}\: g_i(\xb)\overline{\ub}_i\leq\vb_i\leq g_i(\xb)\underline{\ub}_i \quad \text{if}\: g_i(\xb)\geq 0 \: \text{on}\: \Xc_j
\biggr\}, \nonumber
\end{align}
which allows to get rid of the troublesome conditional equality constraints that are now implicitly contained in the inequality constraints. Based on this specification, we can also state a crucial result regarding convexity of $\Zc_j$. 
\begin{lem}\label{lem:convex_Zj}
    Let Assumption~\ref{assum:characterization} hold. Then, every subset $\Zc_j$ as in \eqref{eq:Z_without_eq} is convex.
\end{lem} 
\begin{pf}
    Since every $\Xc_j$ is assumed to be convex, we only need to show convexity of the remaining inequality constraints for the separate cases $g_i(\xb) \geq 0$ and $g_i(\xb) \leq 0$. In the first case, the inequalities can be rearranged as
\begin{align*}
    \underline{\ub}_i g_i(\xb)-\vb_i &\leq 0 \\
    -\overline{\ub}_i g_i(\xb)+\vb_i &\leq 0,
\end{align*}
where the l.h.s of each inequality is a sum of convex functions in $(\xb, \vb)$, since $g_i(\xb)$ is concave for $g_i(\xb)\geq 0$ and both $\underline{\ub}_i$ as well as $ -\overline{\ub}_i$ are negative constants, and $\vb_i$ is affine, i.e., both convex and concave. The same reasoning also applies for the second case
\begin{align*}
    \overline{\ub}_i g_i(\xb)-\vb_i &\leq 0 \\
    -\underline{\ub}_i g_i(\xb)+\vb_i &\leq 0,
\end{align*}
since $g_i(\xb)$ is convex for $g_i(\xb)\leq 0$ and both $\overline{\ub}_i$ as well as $ -\underline{\ub}_i$ are negative constants.
\end{pf}

While the set $\Zc$ itself is still typically nonconvex (except for $\Zc=\Zc_j$), it can be decomposed as the union of a finite number of convex subsets based on Lemma~\ref{lem:convex_Zj}. This result will lead to the specification and solution of convex subproblems later shown in Section~\ref{sec:scenario_based_eval}.

We will now briefly discuss some difficulties encountered when trying to extend these results to the case, where 
$$
    \Gb(\xb) = 
    \begin{pmatrix}
        \Gb_{11}(\xb) & \hdots & \Gb_{1m}(\xb) \\
        \vdots & \ddots & \vdots \\
        \Gb_{m1}(\xb) & \hdots & \Gb_{mm}(\xb)
    \end{pmatrix}
$$
is not diagonal. Assuming for simplicity that $\Gb(\xb)$ is nonsingular, the transformed inequality constraints in \eqref{eq:ineq_constrain} take the form
\begin{alignat}{3}
    && \underline{\ub} &\leq  \Gb^{-1}(\xb) \vb \leq  \overline{\ub} \label{eq:G_nondiag_inequality_all}\\
    &\iff& \underline{\ub}_i &\leq \sum_{j=1}^m \Gb_{ij}^{-1}(\xb) \vb_j \leq  \overline{\ub}_i \quad \forall i \in \{1, ...m\} \label{eq:G_nondiag_inequality_single}
\end{alignat}
where $\Gb_{ij}^{-1}(\xb)$ denotes elements of $\Gb^{-1}(\xb)$ (and not the reciproces of $\Gb_{ij}(\xb)$). A rearrangement of \eqref{eq:G_nondiag_inequality_all} by premultiplying $\Gb(\xb)$ is not viable, since the resulting inequalities are generally not equivalent.
However, in principle, convexity of $\Zc_j$ in the case of diagonal $\Gb(\xb)$ does not depend on this rearrangement, which only served as a simple way of showing that (due to our assumptions on $g_i(\xb)$) every
$$
h_i(\xb, \vb) = \frac{\vb_i}{g_i(\xb)}
$$ 
is quasiconvex whenever $h_i(\xb, \vb)\geq 0$ and quasiconcave whenever $h_i(\xb, \vb)\leq 0$. The sublevel and superlevel sets of $h_i(\xb, \vb)$ involved in specifying each $\Zc_j$ are hence always convex \citep{Boyd2004}. In future work, a similar view on \eqref{eq:G_nondiag_inequality_single} could be useful for determining conditions on $\Gb^{-1}_{ij}(\xb)$ that imply convexity of sets $\Zc_j$. 

\subsection{Choosing the stage cost and terminal ingredients}

The design of stage costs in MPC is an important matter, since they not only act as a performance index by penalizing undesirable system behavior, but their choice affects both the convexity of the resulting OCPs and the stabilization of the control loop. Since the last point also depends on the design of terminal ingredients, we will also discuss those in the following. While our approach will be to design $\hat\ell(\xb, \vb)$ to guarantee certain properties of \eqref{eq:still_NMPC}, its choice will directly influence $\ell(\xb, \ub)$ according to Definition~\ref{def:Rel_vu}, which also allows for straightforward interpretation of the relation between the two. As apparent from \eqref{eq:standard_FL}, the artificial input $\vb=\Gb(\xb)\ub$ represents the actual effect that an input $\ub$ has on the system. Therefore, a stage cost $\hat\ell(\xb,\vb=\Gb(\xb)\ub)$ does not penalize the control effort given by $\ub$, but the actual effect that this effort has on the system. In particular, for states where certain inputs barely affect the system, these specific inputs become less expensive with respect to the cost function, causing the controller to act more aggressively. Whether the use of such a stage cost as a performance index is desirable at all depends very much on the given system.

To retrieve convex subproblems from \eqref{eq:still_NMPC}, we choose $\hat\ell(\xb,\vb)$ convex on every $\Zc_j$. When the goal of the MPC scheme is to stabilize the closed-loop system at the origin, the stage costs are typically chosen positive definite, i.e.
\begin{align*}
    \ell(\xb, \ub) &= 0\quad &&\text{for}\: (\xb, \ub) = (\zerob, \zerob) \\
    \ell(\xb, \ub) &> 0 \quad &&\text{otherwise}.
\end{align*}
As shown in \cite{Klaedtke2022}, positive definiteness of $\hat\ell(\xb,\vb)$ is indeed inherited by $\ell(\xb,\ub)$ for single-input systems with $g_1(\xb=\zerob)\neq 0$ and the same reasoning used there can be applied for the multi-input systems considered here, but with the additional assumption that $g_i(\xb=\zerob)\neq 0$ for all $i \in\{1, ..., m\}$. However, to show the more general case as in Assumption~\ref{assum:characterization}, we will formally restrict the relevant domain of $\ell(\xb,\ub)$.

\begin{lem}\label{lem:posdef_ell}
    Let the relation between $\ub$ and $\vb$ be given by Definition~\ref{def:Rel_vu}, and $\hat\ell(\xb,\vb)$ chosen positive definite on every $\Zc_j$.
    Then, $\ell(\xb,\ub)$ is positive definite for all $(\xb,\ub)\in\Xc \times \tilde\Uc(\xb)$ with
    $$
        \tilde\Uc(\xb) := \left\{\ub\in\Uc \left| \ub_{\Sc(\xb=\zerob)}= \zerob \right\}\right. . 
    $$
\end{lem}
\begin{pf}
    The mapping from $\ub$ to $\vb$ is simply given by $\vb=\Gb(\xb)\ub$. Thus we have $\ell(\xb,\ub)=\hat\ell(\xb,\vb=\Gb(\xb)\ub)$ which is zero iff both $\xb=0$ and $\Gb(\zerob)\ub=\zerob$, and strictly positive otherwise. Now $\Gb(\zerob)$ has a nontrivial nullspace whenever it is singular, leading to the undesirable case that some $\ub\neq \zerob$ are mapped to $\vb=\zerob$ and thus associated with zero cost via $\ell(\xb,\ub)$. The specification of $\tilde\Uc(\xb)$ simply rules out this case by artificially restricting $\ub$ to the trivial nullspace if $\xb=\zerob$ and $\Gb(\zerob)$ is singular.
\end{pf}
Note that this lemma does not further restrict our choice of inputs, but rather shows a result of positive definiteness that also holds for the choice $\ub_{\Sc(\xb)}=\zerob$, which we have already made below Definition~\ref{def:Rel_vu}. In Lemma~\ref{lem:posdef_ell}, the restriction of the domain of $\ell(\xb,\ub)$ amounts to choosing $\ub_{\Sc(\xb=0)} = \zerob$, which is obviously included in $\ub_{\Sc(\xb)} = \zerob$.

 For closed-loop stabilization of the origin, the terminal ingredients $\varphi, \Tc$ are typically chosen so that the closed-loop cost of the controlled system is a lyapunov function, enforcing stability guarantees \citep{Mayne2000}. In \cite{Klaedtke2022}, common choices of $\ell, \varphi, \Tc$ for linear systems were applied to single-input systems \eqref{eq:FL_system}, so we refer to that work for more details. Exactly the same approach can be applied to multi-input systems as long as $\zerob\in\mathrm{int}(\Xc_j)$ holds for one $j\in\{1, ..., s\}$, which we explicitly did not assume in Assumption~\ref{assum:characterization} so as to not restrict the class of applicable systems even further. Since the individual zero level sets $\{\xb\in\Xc|g_i(\xb)=0\}$ are natural boundaries for the partitioning in \eqref{eq:Xpartition}, requiring such a partition to contain the origin in its strict interior would otherwise immediately exclude all systems where $g_i(\zerob)=0$ for any $i\in\{1, ..., m\}$. The existence of such a partition with $\zerob\in\mathrm{int}(\Xc_j)$ can be used to construct a convex and finitely determined terminal set $\Tc$ with nonempty interior, where finite determination is typically lost otherwise. Due to these problems, we will not give a generally applicable choice of $\Tc, \varphi$ for the current scheme but rather show two examples in Section~\ref{sec:examples}, where the previous approach of \cite{Klaedtke2022} can still be applied.

\subsection{Scenario-based evaluation}\label{sec:scenario_based_eval}

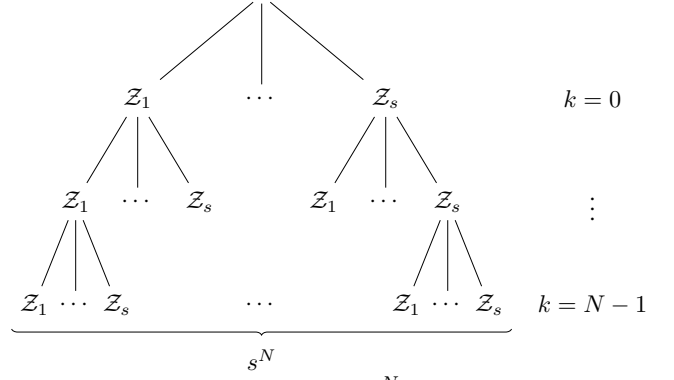
\begin{figure}
    \centering
    \resizebox{0.95\linewidth}{!}{\begin{tikzpicture}[level/.style={sibling distance=18mm/#1}]
\node  (z){}
    child {node  {$\Zc_1$}
        child {node  {$\Zc_1$}
            child {node (a)  {$\Zc_1$}}
            child {node   {$\cdots$}}
            child {node (b)   {$\Zc_s$}}
        }
        child {node   {$\cdots$}}
        child {node   {$\Zc_s$}}
    }
    child {node   {$\cdots$}}
    child {node   {$\Zc_s$}
        child {node  {$\Zc_1$}}
        child {node   {$\cdots$}}
        child {node   {$\Zc_s$}
            child {node (c)  {$\Zc_1$}}
            child {node   {$\cdots$}}
            child {node (d)   {$\Zc_s$}
                child [grow=right] {node (q) {$k = N-1$} edge from parent[draw=none]
                child [grow=up] {node (r) {$\vdots$} edge from parent[draw=none]
                child [grow=up] {node (t) {$k = 0$} edge from parent[draw=none]
                  child [grow=up] {node (u) {} edge from parent[draw=none]}
                }
              }
            }
            }
        }
    };
\path (b) -- (c) node (x) [midway] {$\cdots$}
  child [grow=down] {
    node (y) {}
    edge from parent[draw=none]
  };
 \draw[decoration={brace,mirror,raise=10pt},decorate]
  (a.west) -- node[below=15pt] {$s^N$} (d.east);
\end{tikzpicture}}
    \vspace{-17mm}
    \caption{A tree visualizing the $s^N$ different constraint scenarios, where in each time step $k$ one of the constraints $\begin{pmatrix}\hat\xb(k)^\top &\hat\vb(k)^\top\end{pmatrix}^\top\in\Zc_j$ must hold.}
    \label{fig:scenario_tree}
\end{figure}
Due to our construction, every $(\hat\xb(i), \hat\ub(i))\in\Xc\times \Uc$ has a unique counterpart $\begin{pmatrix}\hat\xb(i)^\top &\hat\vb(i)^\top\end{pmatrix}^\top\in\Zc$ and vice versa, where both are associated with the same stage cost $\ell(\hat\xb(i), \hat\ub(i)) = \hat\ell(\hat\xb(i), \hat\vb(i))$ and lead to the same subsequent state $\hat\xb(i+1) = \Ab\hat\xb(i)+\Bb \Gb(\hat\xb(i)) \hat\ub(i)=\Ab\hat\xb(i)+\Bb \hat\vb(i)$. Therefore, both feasibility and optimality remain unchanged between OCPs \eqref{eq:NMPC} and \eqref{eq:still_NMPC}. Although in \eqref{eq:still_NMPC} the nonlinear equality constraints were replaced with linear ones, and stage and terminal costs $\ell, \varphi$ as well as the terminal constraint set $\Tc$ can simply be chosen convex, the OCP is typically still a nonconvex problem because of the nonconvex constraint set $\Zc$. However, since the condition 
$$
    \begin{pmatrix}
    \hat{\xb}(k)\\
    \hat{\vb}(k)
    \end{pmatrix}\!  \in \Zc =
     \bigcup_{j=1}^s \Zc_j
$$
for each prediction step $k \in\{0, ..., N-1\}$ can be equivalently expressed as 
$$
    \begin{pmatrix}
    \hat{\xb}(k)\\
    \hat{\vb}(k)
    \end{pmatrix}\!  \in \Zc_1 \quad
    \text{or} \quad
    ... \quad
    \text{or} \quad \begin{pmatrix}
    \hat{\xb}(k)\\
    \hat{\vb}(k)
    \end{pmatrix}\!  \in \Zc_s, 
$$
we can split \eqref{eq:still_NMPC} into $s^N$ subproblems, which are indeed convex due to the convexity of every partition $\Zc_j$. To visualize this, consider the tree in Figure~\ref{fig:scenario_tree}, where we will refer to every sequence of constraint sets that is represented by a unique path from the root to one of the leaves as a constraint scenario $\mu\in\{1, ..., s^N\}$. Every such sequence can be compactly represented as 
$$
    \begin{pmatrix}
    \hat{\xb}(k)\\
    \hat{\vb}(k)
    \end{pmatrix}\!  \in \Zc_{\varepsilon_k(\mu)}
$$
with unique coefficients $\varepsilon_0(\mu), ..., \varepsilon_{N-1}(\mu) \in \{1, ..., s\}$ and
$$
    \mu = 1+\sum_{k=0}^{N-1} (\varepsilon_{k}(\mu)-1) s^k.
$$
The solution to \eqref{eq:still_NMPC} can then be obtained by solving the convex subproblems
\begin{align}
\label{eq:convex_subproblem}
V^{(\mu)}(\xb):= \!\!\!\! \min_{\substack{\hat{\xb}(0),\dots,\hat{\xb}(N),\\\hat{\vb}(0),\dots,\hat{\vb}(N-1)}} \!\!\!\!\!\!\!\!\!\!&\,\,\,\,\,\,\,\,\,\,  \varphi(\hat{\xb}(N))+\sum_{i=0}^{N-1} \hat\ell(\hat{\xb}(i),\hat{\vb}(i))\!\!\!\!\!\!\!\!\!\!\!\!\!\span  \\
\nonumber
\text{s.t.} \qquad   \hat{\xb}(0)&=\xb, \\
\nonumber
\hat{\xb}(i+1)&=\Ab\hat\xb(i)+\Bb\hat\vb(i) &\forall i \in \{0,\dots,N\!-1\}, \\
 \nonumber
 \begin{pmatrix}
     \hat{\xb}(i) \\
     \hat{\vb}(i)
 \end{pmatrix} & \in \Zc_{\varepsilon_k(\mu)} &\forall i \in \{0,\dots,N-1\},\\
 \hat{\xb}(N) &\in \Tc\nonumber,
\end{align}
 and determining the optimal constraint scenario
$$
    \mu^\ast := \arg \min_{\mu}\, V^{(\mu)}(\xb).
$$
Note that the feasibility of \eqref{eq:still_NMPC} is equivalent to at least one of the subproblems being feasible.
Finally, we obtain the actual optimal input $\hat\ub^\ast(0)$ by transforming the optimal artificial input of the optimal constraint scenario via the relation in Definition~\ref{def:Rel_vu}. 

For more details regarding the computational complexity of evaluating $s^N$ such scenarios, we refer to considerations made in earlier work \citep{Klaedtke2022}, which do also apply here. Furthermore, note that the general approach of enumerating and evaluating these subproblems is very similar to a common approach for hybrid or piecewise-affine systems thoroughly described in \cite{borrelli_bemporad_morari_2017}. 
In fact, both approaches are so deeply connected that some results \cite[Chp.~17.2]{borrelli_bemporad_morari_2017} can be directly applied to systems \eqref{eq:standard_FL} with (piecewise) affine functions $g_i(\xb)$, i.e., (piecewise) bilinear systems as considered in Example~\ref{exmp:bilinear}. 

\section{Numerical Examples}\label{sec:examples}

We present two numerical examples, where the first is a combination of two single-input examples from \cite{Klaedtke2022} and the second is an example of a bilinear rank-one system from \cite{GHOSH2016}, which we consider in its original single-input form but could easily extend the results to a multi-input variant.
For both examples, we choose $N=15$ and simply consider quadratic stage costs 
$$
    \hat\ell(\xb, \vb) = \xb^\top\Qb\xb+\vb^\top\Rb\vb
$$
with positive definite $\Rb$ and positive semidefinite $\Qb$ of appropriate dimensions, and quadratic terminal cost $\varphi(\xb)=\xb^\top\Pb\xb$, where $\Pb$ is the solution of the discrete algebraic Riccati equation 
$$
\Ab^\top \!\Big( \Pb- \Pb\,\Bb\,\big(\Rb+\Bb^\top \Pb\,\Bb \big)^{-1} \Bb^\top \Pb\Big)\,\Ab - \Pb + \Qb = \zerob.
$$

\begin{exmp}\label{exmp:multi-input}
\begin{figure}[tp]    
        \centering
        \includegraphics[trim=0.3cm 0cm 0.8cm 0.2cm,clip=true]{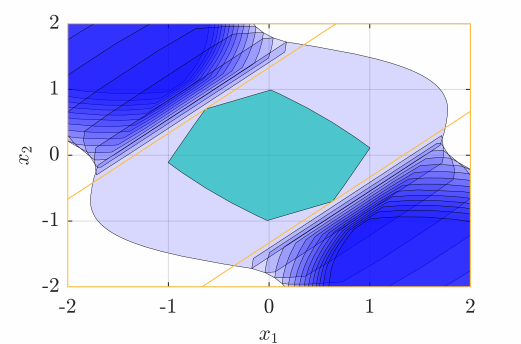}
        \caption{Illustration of the (partially overlapping) feasible sets $\Fc_\mu$ (in blue) and the terminal set $\Tc$ (in cyan) for Example~\ref{exmp:multi-input}. The boundary of the sets $\Xc_j$ is shown in orange.}
        \label{fig:sets_MIMO}
            \end{figure}
            
         \begin{figure}[tp] 
        \centering
        \includegraphics[trim=6cm 10.1cm 6.4cm 10cm,clip=true]{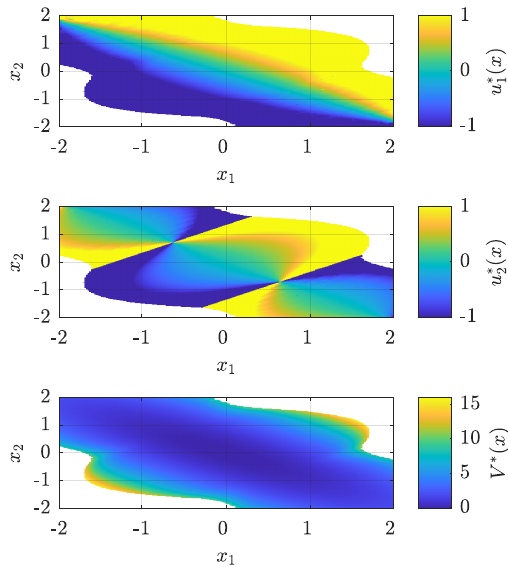}
        \caption{Illustration of the optimal control law $\ub^\ast(\xb)$ (top) and the optimal value function $V^\ast(\xb)$ (bottom) for Example~\ref{exmp:multi-input}.}
        \label{fig:law_cost_MIMO}
    \end{figure}
    Consider a system \eqref{eq:standard_FL} with 
    $$
        \Ab:=\begin{pmatrix}
        1.0 &\, 0.1 \\
        0.1 &\, 1.0 
    \end{pmatrix} \quad \text{and} \quad \Bb:=\begin{pmatrix}
        0.01 & -0.05 \\
        0.05 & -0.01 
    \end{pmatrix},
    $$
    \begin{align}
        g_1(\xb) &= \frac{3}{64}\xb_1^2-\frac{1}{8}\xb_1\xb_2+\frac{3}{64}\xb_2^2-2, \nonumber\\
        g_2(\xb) &= 4\cos\left(\frac{3\pi}{8}(\xb_1-\xb_2)\right), \nonumber
    \end{align}
    which is subject to constraints 
    $$
\Xc:=\{ \xb \in \R^2 \,|\, \|\xb\|_\infty \leq 2 \} \; \text{and} \;
\Uc:=\{ \ub \in \R^2 \,|\, \|\ub \|_\infty \leq 1 \}.
$$
This system can be viewed as a simple combination of Example 1 and 2 in \cite{Klaedtke2022}, where $\Bb$ was changed to allow for more interesting control and $\Uc$ was made individually more restrictive to account for the additional degree of freedom. The way in which \eqref{eq:FL_system} was introduced is equivalent to feedback linearization in \cite{Klaedtke2022} with the parameters $b_0 = 1$ and $\beta = 1$, hence we choose
$$
    \Qb = \begin{pmatrix}
        0.05 & 0.00 \\
        0.00 & 0.05
    \end{pmatrix} \quad \text{and} \quad
    \Rb = \begin{pmatrix}
        0.01 & 0.00 \\
        0.00 & 0.01
    \end{pmatrix}
$$
    to yield comparable results. While $g_1(\xb)$ is convex and non-positive on all of $\Xc$, $g_2(\xb)$ requires a partitioning of $\Xc$, since it is concave and non-negative on the subset 
    $$
        \Xc_1 = \left\{\xb\in \Xc \left|-\frac{4}{3}\leq \xb_1-\xb_2 \leq \frac{4}{3}\right.\right\}
    $$
    but convex and non-positive on the subsets 
    \begin{align*}
        \Xc_2 &= \left\{\xb\in \Xc \left|-4\leq \xb_1-\xb_2 \leq -\frac{4}{3}\right.\right\} \quad \text{and} \\
        \Xc_3 &= \left\{\xb\in \Xc \left|\frac{4}{3}\leq \xb_1-\xb_2 \leq 4\right.\right\}.
    \end{align*}
    To create a partitioning of $\Xc$ which satisfies Assumption~\ref{assum:characterization}, we can generally first consider each $g_i(\xb)$ individually and then construct the intersection of the corresponding partitions. However, since the intersections of $\Xc_1, \Xc_2, \Xc_3$ with $\Xc$ again yield $\Xc_1, \Xc_2, \Xc_3$, this is not technically necessary in this example. Note that since $\zerob \in \mathrm{int}(\Xc_1)$, we can use the method described in \cite{Klaedtke2022} to construct a convex and positively invariant terminal set 
    $$
\Tc \!= 
    \left\{\xb\in\R^n \left|\begin{pmatrix}
\Ib \\
\kappab_{\text{LQR}}
\end{pmatrix}\! \big(\Ab + \Bb \kappab_{\text{LQR}}^\top \big)^{k} \xb \in \Zc_1, \,\forall k \in \N\right.\right\}\!,
$$
which satisfies the system constraints under the linear quadratic regulator (LQR) control law
$$
    \vb(\xb) = -\big(\Rb+\Bb^\top \Pb\,\Bb \big)^{-1} \Bb^\top \Pb\,\Ab\xb := \kappab_{\text{LQR}}^\top\xb.
$$
Because the chosen $\hat\ell, \varphi$ and $\Tc$ satisfy the axioms in \cite{Mayne2000}, we can infer closed-loop stability of the NMPC scheme for this example.

Since we chose $N=15$, we technically need to evaluate $s^N=3^{15}\approx 14.3\cdot 10^6$ constraint scenarios in every time step $k$. However, by solving the corresponding convex feasibility problems \cite[Chp.~11.4]{Boyd2004} associated with each subproblem, we can check in advance, whether a given scenario is feasible for any $\xb\in\Xc$ and disregard all infeasible ones during closed-loop evaluation. Furthermore, using recursive inspection of the scenarios as described in \cite{Klaedtke2022} effectively allows to determine infeasibility of full subtrees in Figure~\ref{fig:scenario_tree}, making this offline preparation tractable. For this example, it turns out that only 31 of the subproblems are feasible and an inner approximation of their respective feasible sets as well as the constructed terminal set are shown in Figure~\ref{fig:sets_MIMO}. Note that the (nonconvex) feasible set of the nonconvex OCP \eqref{eq:still_NMPC} is exactly represented by the union of (convex) feasible sets of all convex subproblems \eqref{eq:convex_subproblem}. Figure~\ref{fig:law_cost_MIMO} shows the optimal control law and optimal costs resulting from the presented scheme and was constructed by sampling the constrained state-space $\Xc$ and solving the subproblems as described in Section~\ref{sec:scenario_based_eval}. Note that the resulting control law is generally not continuous, since the nonconvex OCP \eqref{eq:still_NMPC} does not satisfy the continuity conditions stated in \cite{borrelli_bemporad_morari_2017}, and the most noticeable examples of discontinuity can be seen at the boundaries of $\Xc_i$, where $g_2(\xb)$ changes signs. 
    \end{exmp}
    \begin{exmp} \label{exmp:bilinear}
    \begin{figure}[tp] 
        \centering
        \includegraphics[trim=5.9cm 10.1cm 6.4cm 10cm,clip=true]{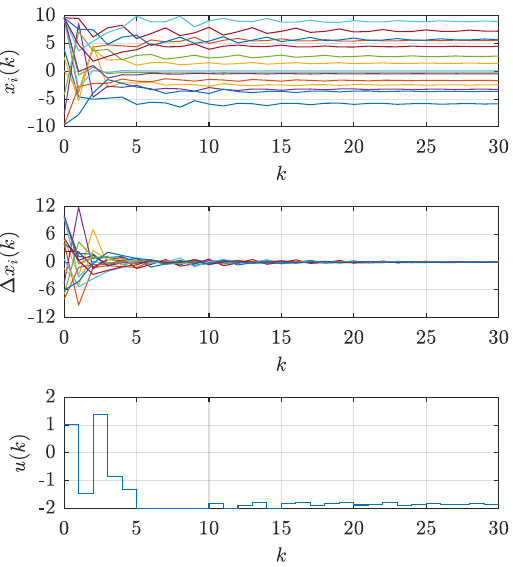}
        \caption{Closed-loop simulation of the bilinear network considered in Example~\ref{exmp:bilinear} showing the system trajectory in original (top) and shifted (middle) coordinates as well as the applied input (bottom).}
        \label{fig:closed_loop_network}
    \end{figure}
        Consider a bilinear single-input system
        $$
            \xb(k+1) = \Ab\xb(k)+ \Nb\xb(k) u(k),
        $$
        where $\Nb$ is rank one and hence can be decomposed as $\Nb=\bb \cb^\top$. In general, bilinear systems are very useful to describe networks with nodes $\xb_i$, where the entries of $\Ab$ represent the usual (uncontrolled) edges and $\Nb$ represents edges, which are dependant on the current input, i.e., $u(k)$ strengthens or weakens these connections described by $\Nb$. Typical applications are given by \cite{GHOSH2016} as traffic or biochemical networks. The specific system we consider is an example from \cite{GHOSH2017} with $n=15$ nodes and general structure given by the graph shown in Figure~1 of \cite{GHOSH2017}, which determines the configuration of zero and non-zero entries in $\Ab, \Nb$, while the values of non-zero entries themselves are randomly chosen. Even though an extension to the multi-input case with individual rank-one matrices $\Nb_i$ would be straightforward for our scheme, we stick with the original example and can easily see that it satisfies the required structure of \eqref{eq:standard_FL} with $\Bb=\bb$ and $\Gb(\xb) = g_1(\xb) = \cb^\top\xb$. We consider the constraint sets
            $$
        \Xc = \left\{\xb\in \R^n \left|\;\|\xb \|_\infty \leq 10 \, \right.\right\} \; \text{and} \;
        \Uc = \left\{u \in \R \left|\; |u| \leq 2 \right.\right\},
        $$
        and the hyperplane given by $g_1(\xb) = \cb^\top\xb=0$ naturally splits the constrained state-space $\Xc$ into the two partitions
        \begin{equation}
\label{eq:constraints_bilinear_ex}
        \Xc_1 = \left\{\xb\in \Xc \left| g_1(\xb)\!\leq 0 \!\right.\right\}, 
        \Xc_2 = \left\{\xb\in \Xc \left| g_1(\xb)\! \geq 0 \!\right.\right\}.
        \end{equation}
         Note that, while in general this class of bilinear systems is very similar to the one investigated in \cite{NMPC2021Exact}, we here have $g_1(\zerob) = 0$, which is an important distinction because now the origin lies on the boundary of the two partitions and hence is not contained in either of their interiors. 
         Since the input becomes powerless near the origin, i.e., strengthening or weakening edges of the network is meaningless if there is nothing to transfer between the nodes, we will instead aim to stabilize the system at a different equilibrium $(\xb^\circ, u^\circ)\neq(\zerob,0)$. For the sake of this example, we randomly choose a pair $(\xb^\circ, u^\circ)\in\Xc\times\Uc$, which satisfies
         $$
            \xb^\circ = \Ab \xb^\circ + \Nb \xb^\circ u^\circ \quad \text{and} \quad g_1(\xb^\circ)\neq 0,
         $$
         and introduce the shifted coordinates
         \begin{equation}
             \Delta\xb = \xb-\xb^\circ \quad \text{and}\quad \Delta u = u-u^\circ. 
             \label{eq:delta_transform}
         \end{equation}
         It is easy to see that the system dynamics in these new coordinates are given by
         \begin{align}
             \Delta\xb(k+1) &= (\Ab+\Nb u^\circ)\Delta\xb(k)+\bb\cb^\top(\Delta\xb(k)+\xb^\circ)\Delta u(k)\nonumber \\
             &= \tilde\Ab \Delta\xb(k)+\bb\tilde g(\Delta\xb(k)) \Delta u(k), \label{eq:delta_system}
         \end{align}
        which is a bilinear system system satisfying the structure assumed in \cite{NMPC2021Exact,Klaedtke2022}, since we now have $\tilde g(\Delta\xb=\zerob)\neq 0$.
        Shifting the constraint sets \eqref{eq:constraints_bilinear_ex} in accordance with the coordinate transformation \eqref{eq:delta_transform}, we can now consider the OCP \eqref{eq:still_NMPC} for the artificial input $\Delta v = \tilde g(\Delta \xb) \Delta u$ and shifted state $\Delta\xb$. Choosing simple quadratic stage costs with $\Qb = \Ib_n$ and $\Rb = 1$, we can construct a terminal set the same way as in Example~\ref{exmp:multi-input}. Note that since the sets $\Zc_j$ in this case are polyhedral, the constructed terminal set $\Tc$ is polyhedral as well \citep{Gilbert1991} and having both polyhedral constraints and quadratic costs leads to the convex subproblems \eqref{eq:convex_subproblem} being reframable as quadratic programs (QPs), for the solution of which many well-known algorithms exist (see, e.g., \cite{Boyd2004}). For stabilization of the arbitrarily chosen equilibrium only $8$ of the $s^N=2^{15}=32768$ constraint scenario turn out to be feasible, which might however drastically change for different system parameters and a different equilibrium. The Figure~\ref{fig:closed_loop_network} shows a closed-loop simulation of the controlled network for an initial state $\xb(\zerob)$ randomly chosen in the feasible set of \eqref{eq:still_NMPC}. As expected, the system is being stabilized at the chosen equilibrium while adhering to the state and input constraints.  
    \end{exmp}
\section{Summary and Outlook}
    We showed that a special class of (nonconvex) NMPC problems admits an exact solution by considering a finite number of convex subproblems with suitable stage costs. These results are applicable to a special class of nonlinear discrete-time systems, where the main contribution of this work lies in extending the results of \cite{Klaedtke2022} to systems with multiple inputs and removing a technical assumption, which now allows the scheme to be applied, among others, to a class of rank-one bilinear networks considered in \cite{GHOSH2016}. Both aspects were demonstrated 
    with numerical examples.

    The results of this work are another step towards using the scheme for actual practical problems and gradually moving away from purely numerical examples. Although the class of applicable systems is continuously being extended, it still remains very restrictive and future work should aim at further extending this class while also addressing the drawbacks of the current scheme. As discussed in Section~\ref{sec:Constraint_Trafo}, the extension to nondiagonal $\Gb(\xb)$ does present difficulties, but should nevertheless be studied in more detail in the future because it would significantly increase the class of applicable systems. Furthermore, different choices of stage costs $\hat\ell(\xb,\vb)$ and their effect on $\ell(\xb,\ub)$ should be investigated, since the current choice promotes some undesired discontinuous behavior as shown in Example~\ref{exmp:multi-input} in Section~\ref{sec:examples}. Finally, even though the previous choice of terminal set $\Tc$ made in \cite{Klaedtke2022} is applicable to both examples in Section~\ref{sec:examples}, it is unsuitable for stabilizing systems at equilibria $(\xb^\circ, \ub^\circ)$ with any $g_i(\xb^\circ)=0$. The possibility of other choices for $\Tc$ should hence be investigated for this case.


\begin{thebibliography}{14}
\footnotesize
\setlength{\parskip}{0pt}
\setlength{\itemsep}{1pt plus 0.3ex}
\providecommand{\natexlab}[1]{#1}
\providecommand{\url}[1]{\texttt{#1}}
\providecommand{\urlprefix}{URL }
\expandafter\ifx\csname urlstyle\endcsname\relax
  \providecommand{\doi}[1]{doi:\discretionary{}{}{}#1}\else
  \providecommand{\doi}{doi:\discretionary{}{}{}\begingroup
  \urlstyle{rm}\Url}\fi

\bibitem[{Bacic et~al.(2002)Bacic, Cannon, and Kouvaritakis}]{Bacic2002}
Bacic, M., Cannon, M., and Kouvaritakis, B. (2002).
\newblock Feedback linearization {MPC} for discrete-time bilinear systems.
\newblock In \emph{Proc. of the 15th IFAC World Congress}, 159--164.

\bibitem[{Borrelli et~al.(2017)Borrelli, Bemporad, and
  Morari}]{borrelli_bemporad_morari_2017}
Borrelli, F., Bemporad, A., and Morari, M. (2017).
\newblock \emph{Predictive Control for Linear and Hybrid Systems}.
\newblock Cambridge University Press.
\newblock \doi{10.1017/9781139061759}.

\bibitem[{Boyd and Vandenberghe(2004)}]{Boyd2004}
Boyd, S. and Vandenberghe, L. (2004).
\newblock \emph{Convex Optimization}.
\newblock Cambrige University Press.

\bibitem[{Gao et~al.(2012)Gao, Yang, Wang, and Yu}]{Gao2012}
Gao, D., Yang, Q., Wang, M., and Yu, Y. (2012).
\newblock Feedback linearization optimal control approach for bilinear systems
  in {CSTR} chemical reactor.
\newblock \emph{Intelligent Control and Automation}, 3(3), 274--277.

\bibitem[{Ghosh and Ruths(2016)}]{GHOSH2016}
Ghosh, S. and Ruths, J. (2016).
\newblock Structural control of single-input rank one bilinear systems.
\newblock \emph{Automatica}, 64, 8--17.
\newblock \doi{https://doi.org/10.1016/j.automatica.2015.10.053}.

\bibitem[{Ghosh et~al.(2017)Ghosh, Ruths, and Yeo}]{GHOSH2017}
Ghosh, S., Ruths, J., and Yeo, A. (2017).
\newblock Graphical coprime walk algorithm for structural controllability of
  discrete-time rank-one bilinear systems.
\newblock \emph{Automatica}, 86, 166--173.
\newblock \doi{https://doi.org/10.1016/j.automatica.2017.08.029}.

\bibitem[{Gilbert and Tan(1991)}]{Gilbert1991}
Gilbert, E.G. and Tan, K.T. (1991).
\newblock Linear systems with state and control constraints: The theory and
  application of maximal output admissible sets.
\newblock \emph{IEEE Trans. Autom. Control}, 36(9), 1008--1020.

\bibitem[{Isidori(1995)}]{Isidori1995}
Isidori, A. (1995).
\newblock \emph{Nonlinear Control Systems}.
\newblock Springer.

\bibitem[{Kl{\"a}dtke and Schulze~Darup(2022)}]{Klaedtke2022}
Kl{\"a}dtke, M. and Schulze~Darup, M. (2022).
\newblock Convex reformulations for a special class of nonlinear mpc problems.
\newblock In \emph{2022 European Control Conference}, 761--768.
\newblock \doi{10.23919/ECC55457.2022.9838061}.

\bibitem[{Lautenschlager et~al.(2015)Lautenschlager, Kruppa, and
  Lichtenberg}]{Lautenschlager2015}
Lautenschlager, B., Kruppa, K., and Lichtenberg, G. (2015).
\newblock Convexity properties of the model predictive control problem for
  subclasses of multilinear time-invariant systems.
\newblock In \emph{Proc. of 5th IFAC Conf.~on Nonlinear Model Predictive
  Control}, 148--153.

\bibitem[{Mayne et~al.(2000)Mayne, Rawlings, Rao, and Scokaert}]{Mayne2000}
Mayne, D.Q., Rawlings, J.B., Rao, C., and Scokaert, P.O.M. (2000).
\newblock {Constrained model predictive control: Stability and optimality}.
\newblock \emph{Automatica}, 36, 789--814.

\bibitem[{Rockafellar(1993)}]{Rockafellar1993}
Rockafellar, R.T. (1993).
\newblock Lagrange multipliers and optimality.
\newblock \emph{SIAM Review}, 35(2), 183--238.

\bibitem[{Schulze~Darup et~al.(2021)Schulze~Darup, Klädtke, and
  Mönnigmann}]{NMPC2021Exact}
Schulze~Darup, M., Klädtke, M., and Mönnigmann, M. (2021).
\newblock Exact solution to a special class of nonlinear mpc problems.
\newblock \emph{IFAC-PapersOnLine}, 54(6), 290--295.
\newblock \doi{https://doi.org/10.1016/j.ifacol.2021.08.559}.
\newblock 7th IFAC Conference on Nonlinear Model Predictive Control 2021.

\bibitem[{Soroush and Kravaris(1992)}]{Soroush1992}
Soroush, M. and Kravaris, C. (1992).
\newblock Discrete‐time nonlinear controller synthesis by input/output
  linearization.
\newblock \emph{AIChE Journal}, 38, 1923--1945.

\end{thebibliography}
\end{document}